\documentstyle[aps,prl,preprint,epsfig]{revtex} 
\draft
\newcommand{\pathfigsiiiD}{.}
\newcommand{\pathfigsQ}{.}

\newcommand{\be}{\begin{equation}}
\newcommand{\ee}{\end{equation}}

\begin{document}
\pagestyle{myheadings}
\markboth{Helbing/Hennecke/Treiber: Phase Diagram of Traffic States ...}
{Helbing/Hennecke/Treiber: Phase Diagram of Traffic States ...}



\title{Phase Diagram of Traffic States in
the Presence of Inhomogeneities}
\author{Dirk Helbing$^{1,2}$,
Ansgar Hennecke$^1$,
and Martin Treiber$^1$}
\address{$^1$II. Institute of 
         Theoretical Physics, University of Stuttgart,
         Pfaffenwaldring 57/III, 70550 Stuttgart, Germany\\
$^2$Department of Biological Physics, E\"otv\"os University,
Budapest, Puskin u 5--7, H--1088 Hungary}
\maketitle
\begin{abstract}
We present a phase diagram of the different kinds of congested
traffic that are triggered by disturbances when passing
ramps or other spatial inhomogeneities of a freeway. The simulation results
obtained by the nonlocal, gas-kinetic-based traffic model 
are in good agreement with empirical findings. They 
allow to understand the observed transitions between free and various
kinds of congested traffic, among them localized clusters,
stop-and-go waves, and different types of ``synchronized'' traffic.
We also give analytical conditions for the existence of these states which
suggest that
the phase diagram is universal for a class of different microscopic and
macroscopic traffic models.
\end{abstract}


For some years now,
the various observed nonlinear states of freeway traffic flow,
including traffic jams and stop-and-go waves, 
have attracted the interest
of a rapidly growing community of physicists \cite{TGF95,book}. 
The same applies to simulation approaches that allow to reproduce
the 
transitions between the different collective states of 
motion \cite{book,Schreck,kk-94,Letter}. A particular attention received 
the recent discovery of a first-order phase transition to 
``synchronized'' congested traffic \cite{kr-96b,kr-97}, which
stimulated an intense investigation of the complex phenomena 
associated with ramps, yielding an explanation of the 
hysteretic phase transition \cite{Lee,Letter}.
It is also known from other effectively one-dimensional 
many-body systems with short-range interactions that  
localized inhomogeneities can cause a variety of phenomena, including
phase transitions \cite{phase} and spontaneous symmetry breaking
\cite{sym}. 
\par
To characterize the parameter-dependence of the
possible states of a system resulting in the long run, phase diagrams
are a very powerful method. They are of great
importance in thermodynamics with various applications in 
metallurgy, chemistry, etc. Moreover, they allow to compare very different
kinds of systems like equilibrium and nonequilibrium ones, or microscopic and
macroscopic ones, whose  equivalence 
cannot simply be shown 
by transformation to normal forms \cite{manneville,Collet}. 
Defining universality classes by mathematically equivalent
phase diagrams, one can even classify so different systems as
physical, chemical, biological and social ones, as done in systems theory.
\par
\begin{figure}

\vspace{0\unitlength}

\begin{center}
\hspace*{-6\unitlength}
\includegraphics[width=160\unitlength]{\pathfigsQ/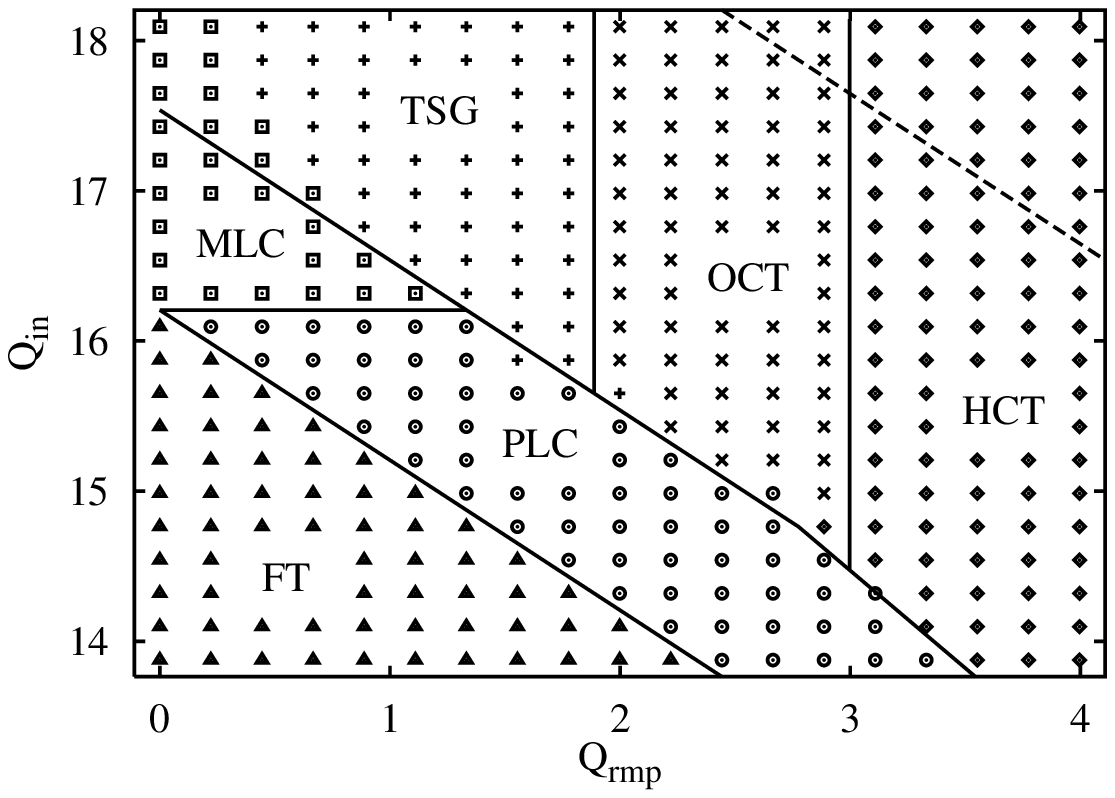} \\[-12mm]
\end{center}

\caption[]{\label{phasedia}
Phase diagram of the  traffic states in the vicinity of
an on-ramp 
as a function of the 
inflows $Q_{\rm in}$ and $Q_{\rm rmp}$ on the main road and the 
on-ramp for fixed ramp length $L=56$. 
The different states are classified at
$t=135$, i.e., after a sufficiently long transient period.
Displayed are homogeneous congested traffic ($\Diamond$),
oscillatory congested traffic ($\times$),
triggered stop-and-go traffic (+),
moving localized clusters ($\Box$),
pinned localized clusters ($\odot$),
and free traffic ($\bigtriangleup$).
The states are triggered by a fully developed localized cluster
travelling upstream and passing the ramp (cf. Fig.
\protect\ref{rho3d}). Solid lines indicate the theoretical phase boundaries.
The dashed line represents the condition
$(Q_{\rm in} + Q_{\rm rmp}) = Q_{\rm max}$ that characterizes the maximum 
downstream flow for
which a (possibly unstable) equilibrium solution exists.
Above this line, extended congestion develops even without any perturbation. 
}
\end{figure}
\begin{figure}
\vspace{-8\unitlength}
\begin{center}
\hspace*{-10\unitlength}   
   \includegraphics[width=80\unitlength]{\pathfigsiiiD/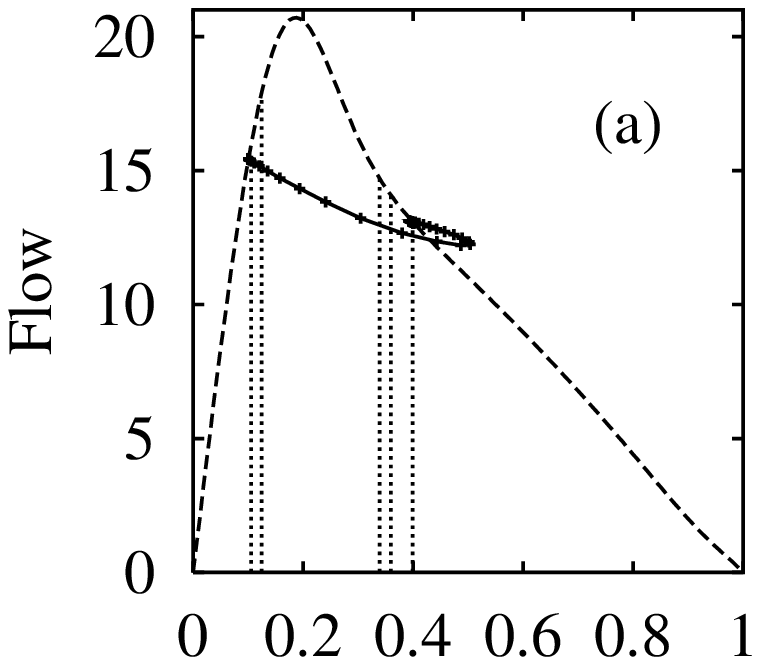}
       \hspace{-15\unitlength}
   \includegraphics[width=80\unitlength]{\pathfigsiiiD/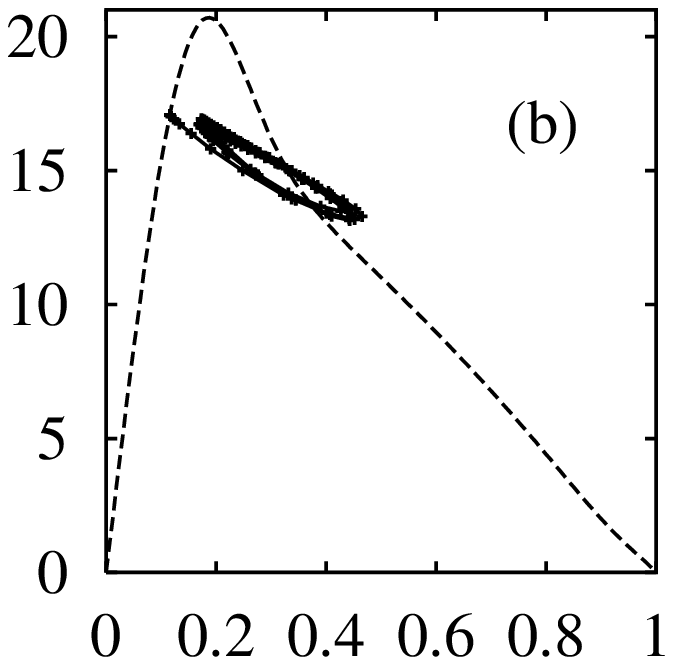} 
      \\[-15\unitlength]
\hspace*{-10\unitlength}
   \includegraphics[width=80\unitlength]{\pathfigsiiiD/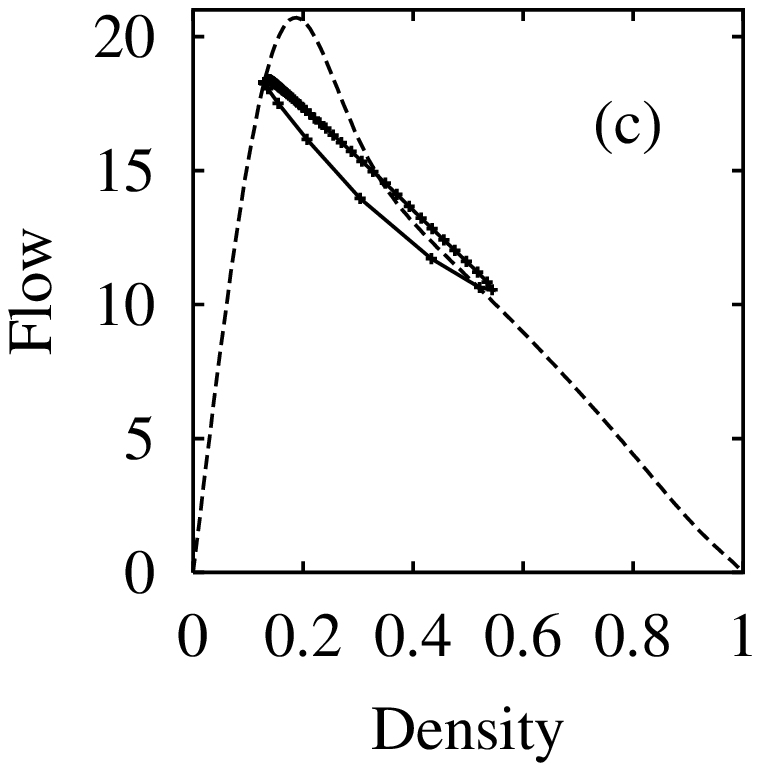}
       \hspace{-15\unitlength}
   \includegraphics[width=80\unitlength]{\pathfigsiiiD/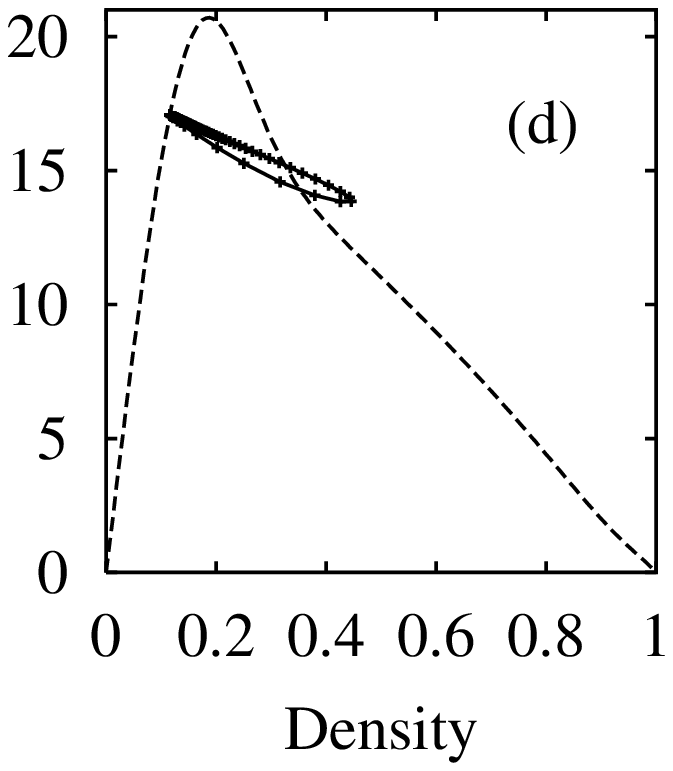} \\[-12mm]

\end{center}

\caption[]{\label{rhoQ}
Dynamics in the flow-density space at the fixed position $x=-560$ upstream
of the ramp. Dashed lines represent 
the flow-density diagram (i.e. the equilibrium flow-density relation),
dotted lines the position of the critical densities $\rho_{{\rm c}i}$. 
(a) Shown is the transition from free
traffic (symbols at the left side of the flow-density diagram) to
homogeneous congested traffic (symbols at the right side).
The other illustrations display (b) oscillatory congested traffic,
(c) triggered stop-and-go traffic, and (d) a moving localized cluster.
}
\end{figure}

In the following, we will present and explain the phase diagram
that we obtained for the nonlocal, gas-kinetic-based traffic model
\cite{Letter}, when we studied a freeway with a ramp 
in the presence of a single perturbation [Fig.~\ref{phasedia}]. 
By systematical variation of the inflow at the upstream freeway
boundary {\em and} the
ramp, we were able to reproduce the observed transitions between different
traffic states, some of which have not been explained before. We will
also present analytical conditions for the existence of the various
traffic states. Since we are interested
in the generic properties of the model, we will formulate it 
in dimensionless quantities \cite{overview}
by measuring times
in units of the relaxation time $\tau \approx 40$\,s,
and distances in units
of the inverse $1/\rho_{\rm max}$ of the maximum vehicle
density $\rho_{\rm max}\approx 140$\,vehicles/km. The equation for the
vehicle density $\rho(x,t)$ at position $x$ and time $t$ is a continuity
equation with a sink/source term and reads
\begin{equation}
\label{eqrhoscal} 
 \frac{\partial \rho}{\partial t} + 
 \frac{\partial (\rho V)}{\partial x} 
 = \frac{Q_{\rm rmp}(x,t)}{L_{\rm rmp}} \, .
\end{equation}
$V$ denotes the average vehicle velocity, $L_{\rm rmp}$ is the length of
the ramp, $Q_{\rm rmp}$ is the flow of vehicles entering the free\-way
($Q_{\rm rmp} > 0$) or leaving it ($Q_{\rm rmp} < 0$) along the ramp
of length $L_{\rm rmp}$,
divided by the number $n$ of freeway lanes. The velocity equation
contains a convection term (due to the movement of the velocity profile
with velocity $V$), a pressure term (reflecting dispersion effects due to
a finite velocity variance $\theta$), 
a relaxation term (describing an adaptation to a desired velocity $V_0$), 
and an interaction term (corresponding to braking maneuvers)
\cite{overview}:
\begin{eqnarray}
\label{eqVscal}
  \frac{\partial V}{\partial t} 
       + V \frac{\partial V}{\partial x}
   &=& - \frac{1}{\rho} \frac{\partial(\rho\theta)}{\partial x}
       + (V_0-V) \nonumber \\
   & & - PA(\rho_{\rm a}) \frac{(\rho_{\rm a} V_{\rm a})^2}
          {(1-\rho_{\rm a})^2} B(\delta_V) \, .
\end{eqnarray}
%
$V_0$ has a meaning similar to the Reynolds number
(since it allows for traffic instabilities when it exceeds the
value $V_{\rm c} = 61$). In addition,
$P$ is a scaled cross section.
\be
\label{Ascal}
A(\rho) = 0.171 + 0.417 \{ \tanh [ 10 (\rho-0.27) ] + 1 \}
\ee
represents a structure factor, which is of order unity and determines the
variance via the constitutive relation $\theta = A(\rho) V^2$, but also the
form of the flow-density relation in equilibrium \cite{overview} 
(see Fig.~\ref{rhoQ}). Finally,
\begin{equation}  
\label{B}
B(\delta_V) =   2 \left[ 
    \delta_V \frac{\mbox{e}^{-\delta_V^2/2}}{\sqrt{2\pi}}
           + (1+\delta_V^2) 
    \int_{-\infty}^{\delta_V} dy \, \frac{\mbox{e}^{-y^2/2}}{\sqrt{2\pi}}
                 \right]
\end{equation}
with the dimensionless velocity difference 
$\delta_V = (V-V_{\rm a})/\sqrt{\theta+\theta_{\rm a}}$ is
a Boltzmann factor arising from the vehicle interactions.
An index ``a'' indicates that the respective quantity is evaluated at
the advanced ``interaction point'' $x_{\rm a} = x + \gamma (1 + T V)$
rather than at the actual position $x$. The related nonlocality reflects
the anticipative driver behavior and is essential for the realistic
properties of the model and its robust and efficient numerical solution.
$\gamma$ is 
an anticipation factor and $T$ 
about the safe time headway.
In our simulations, we used the parameter values $V_0=171$,
$P=0.31$, $\gamma=1.2$, and $T=0.043$, which are typical for
Dutch freeways \cite{overview}, but
our results are not very sensitive to 
their particular choice. 
\par
The nonlocal, gas-kinetic-based traffic 
model reproduces the characteristic properties of traffic flows 
\cite{overview} formulated by Kerner and Rehborn \cite{kr-96a}. 
Furthermore, for a homogeneous freeway {\em without} ramps,
it shows the phase (``instability'') 
diagram \cite{overview} that was postulated
by Kerner and Konh\"auser on the basis of macroscopic simulations \cite{kk-94}:
For a given density $\rho$, there exists a state of homogeneous
traffic with equilibrium velocity $V_{\rm e}(\rho)$ and
equilibrium flow $Q_{\rm e}(\rho) = \rho V_{\rm e}(\rho)$
(see Fig.~\ref{rhoQ}), and there are four critical densities
$\rho_{{\rm c}i}$. For densities 
$\rho < \rho_{\rm c1}$ and $\rho > \rho_{\rm c4}$,
homogeneous traffic is stable with respect to localized perturbations, 
and for a range $\rho_{\rm c2} < \rho < \rho_{\rm c3}$
of intermediate densities, it is 
linearly unstable, giving rise to cascades of traffic jams 
(``stop-and-go traffic''). For the two density regimes 
$\rho_{\rm c1} \le \rho \le \rho_{\rm c2}$ and 
$\rho_{\rm c3} \le \rho \le \rho_{\rm c4}$ between the stable and
the linearly unstable regions, it is metastable, i.e., it behaves
nonlinearly unstable with respect to perturbations exceeding a 
certain critical amplitude, but otherwise stable.
For the discussion of the phase diagram, 
an additional range
$\rho_{\rm cv}  \le \rho \le \rho_{\rm c3}$ of convective stability 
will be important, where homogeneous traffic is linearly unstable, 
but the growing perturbations are convected 
away from any fixed location
\cite{manneville}.
\par
We will denote by $Q_{{\rm c}i}=Q_{\rm e}(\rho_{{\rm c}i})$ 
the equilibrium flows corresponding to the 
critical densities introduced above. 
Moreover, there exists a characteristic outflow $Q_{\rm out}$
from traffic jams, stop-and-go waves, etc.
\cite{overview,kr-96a}, 
that is nearly independent of the surrounding density and the type 
of congested traffic, at least if there is no ramp.
If traffic relaxes at the location of an on-ramp, the observed
outflow is $\tilde{Q}_{\rm out}(Q_{\rm rmp},L_{\rm rmp}) 
\le Q_{\rm out}$, 
but we find
$\tilde{Q}_{\rm out}\to Q_{\rm out}$ for 
$L_{\rm rmp}\to\infty$ or $Q_{\rm rmp}\to 0$
\cite{Letter}. For our model parameters,
we obtain
$\rho_{\rm c1}   = 0.11$, 
$\rho_{\rm c2}   = 0.12$, 
$\rho_{\rm cv}   = 0.34$, 
$\rho_{\rm c3}   = 0.36$, and
$\rho_{\rm c4}   = 0.40$, 
related to
$Q_{\rm c1}  =16.2 $,   
$Q_{\rm c2}  =17.5 $,   
$Q_{\rm cv}=14.6 $,   
$Q_{\rm c3}  =14.1 $, and  
$Q_{\rm c4}  =13.2 $.   
If $L\ge 50$, we obtain $\tilde{Q}_{\rm out} \approx Q_{\rm c2}$,
otherwise $\tilde{Q}_{\rm out}$ decreases with growing $Q_{\rm rmp}$
and decreasing $L_{\rm rmp}$. 
For these values (which, with the exception of $\tilde{Q}_{\rm out}$, are
determined from situations {\em without} ramps!), the equations for the
phase boundaries suggested below agree quantitatively with the 
ones in the phase diagram 
in Fig.~\ref{phasedia}. 
\par
\begin{figure}
\vspace{-8\unitlength}
\begin{center}
   \includegraphics[width=80\unitlength]{\pathfigsiiiD/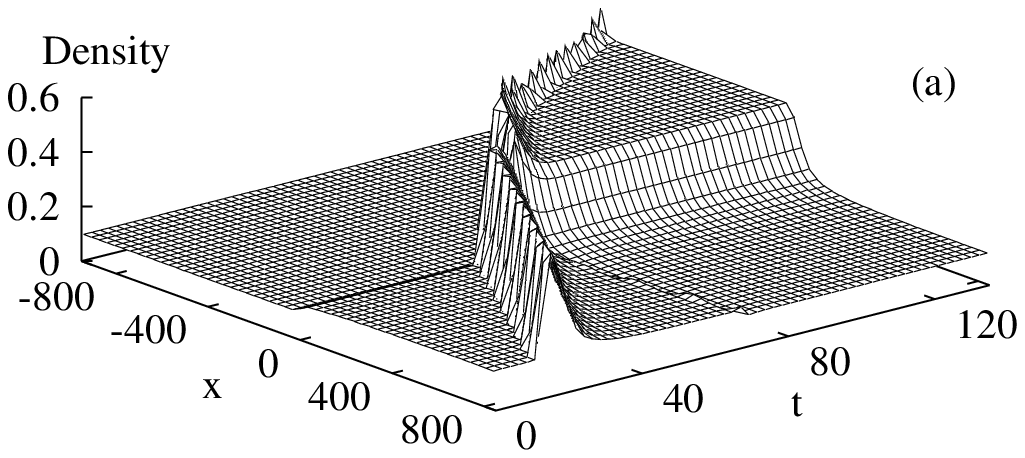}
      \\[-12\unitlength]
   \includegraphics[width=80\unitlength]{\pathfigsiiiD/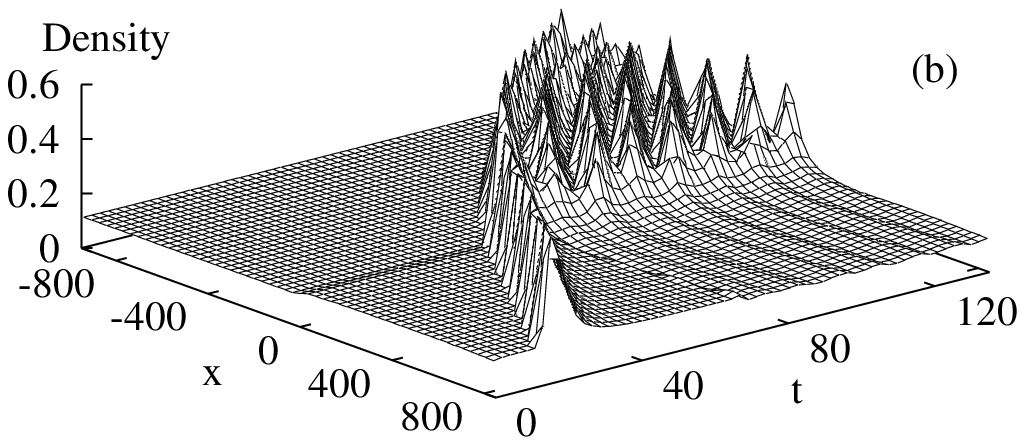}
      \\[-12\unitlength]
   \includegraphics[width=80\unitlength]{\pathfigsiiiD/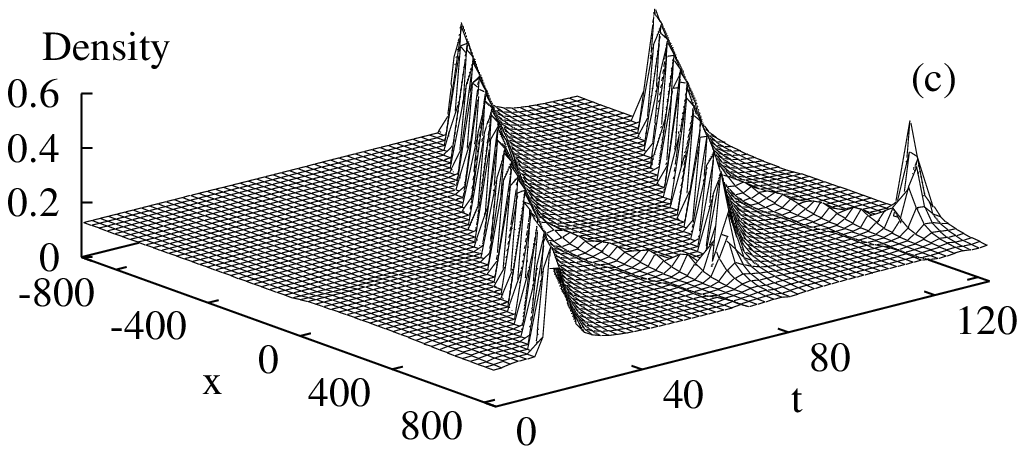} 
      \\[-15\unitlength]
   \includegraphics[width=80\unitlength]{\pathfigsiiiD/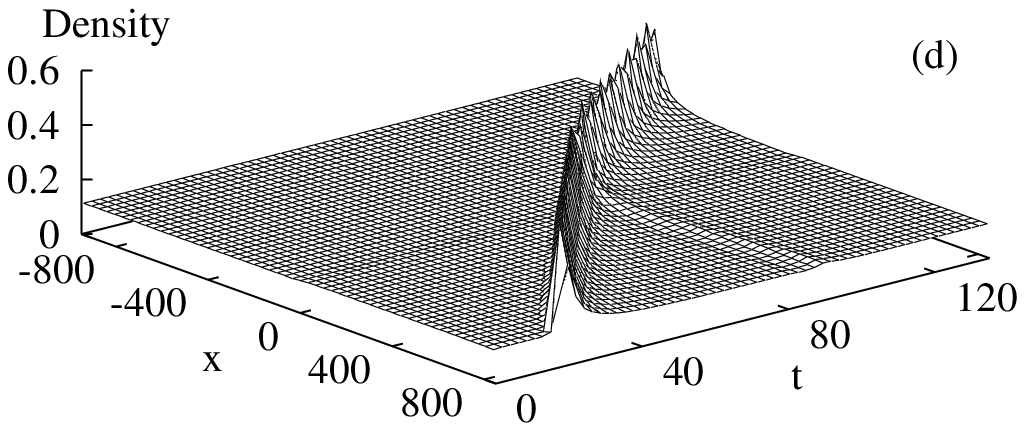}
      \\[-12\unitlength]
\end{center}

\caption[]{\label{rho3d}
Spatiotemporal dynamics of typical representatives of the states
depicted in Fig.~\protect\ref{phasedia}.
The respective states are
(a) homogeneous congested traffic (HCT), (b) oscillatory
congested traffic (OCT), (c) triggered stop-and-go traffic (TSG), and
(d) a moving localized cluster (MLC).
$x=0$ corresponds to the middle of the ramp.
}
\end{figure}

In our simulations, we investigated 
the congested traffic states 
that formed near an
on-ramp when passed by a perturbation on the freeway (Fig.~\ref{rho3d}). 
For the perturbation 
we chose a fully developed density cluster that did not change its amplitude
or shape anymore and travelled upstream with constant speed. 
When the density cluster reached the ramp, it
induced different kinds of congested states
(Figs.~\ref{phasedia} through \ref{rho3d}),
depending
on the inflows $Q_{\rm in}$ and $Q_{\rm rmp}$ at the upstream boundary 
of the freeway and the ramp, respectively.
\par
Now, we will explain the different states in the phase diagram
starting with relatively high ramp flows.
When traffic breaks down (a sufficient criterium
for this is given by $(Q_{\rm out} + Q_{\rm rmp}) > Q_{\rm max}$,
where $Q_{\rm max}$ denotes the maximum of the equilibrium flow),
%
we find the formation of a growing, 
extended region of congested traffic with 
average flow $Q_{\rm cong} = (\tilde{Q}_{\rm out} - Q_{\rm rmp})$ 
in front of the downstream end of the on-ramp \cite{Letter}. 
Because of the fixed downstream end, any perturbation will be
convected out of the congested region, if the homogeneous solution
of density $\rho_{\rm cong}$ and flow 
$Q_{\rm cong} = Q_{\rm e}(\rho_{\rm cong})$ is convectively stable, i.e. 
if $Q_{\rm cong} \le Q_{\rm cv}$.
In this case, we have homogeneous congested traffic 
(HCT) [Fig.~\ref{rho3d}(a)], 
which corresponds to an equilibrium solution on the 
high-density branch of the flow-density diagram [Fig.~\ref{rhoQ}(a)].
Otherwise, oscillatory
congested traffic (OCT) forms [Figs.~\ref{rho3d}(b) and \ref{rhoQ}(b)].
Thus, the boundary between HCT and OCT is given by
\be
\label{IST_HST}
\mbox{HCT-OCT:}\ \  Q_{\rm rmp} = \tilde{Q}_{\rm out} - Q_{\rm cv} \, .
\ee
The oscillation amplitudes grow with decreasing ramp flow, until 
they reach the low-density part of the flow-density diagram 
associated with free traffic [Fig.~\ref{rhoQ}(c)]. This means that we
have an alternation between free and congested traffic which defines
the state of triggered stop-and-go waves (TSG)
[Fig.~\ref{rho3d}(c)]. It turns out,
that TSG is characterized by $Q_{\rm TSG} \ge Q_{\rm TSG}^{\rm min}$,
where $Q_{\rm TSG}$ denotes the average flow of the TSG state. Hence,
the boundary between TSG and OCT is given by
\be
\label{TSG-IST}
\mbox{OCT-TSG:}\ \  Q_{\rm rmp}=\tilde{Q}_{\rm out}-Q_{\rm TSG}^{\rm min}
\, .
\ee
However, the precise value of $Q_{\rm TSG}^{\rm min}$ is hard to determine. 
\par
The minimal inflows to sustain the TSG states can be derived from
the triggering mechanism illustrated in Fig.~\ref{rho3d}(c).
When passing the ramp, the initial localized perturbation causes 
a secondary perturbation travelling downstream \cite{kerner-ramp}.
Since the amplitude of this secondary perturbation is always small,
regardless of the amplitude of the primary perturbation, it only grows
on the instability condition $Q_{\rm down}> Q_{\rm c2}$, where
$Q_{\rm down} = (Q_{\rm in} + Q_{\rm rmp})$ denotes the flow downstream of
the ramp. With growing amplitude, the triggered perturbation
changes its propagation speed, reverses its direction, 
and finally induces another small perturbation
when passing the ramp, etc. In this way, the interplay
of the perturbations with the ramp defines an intrinsic timescale and
wavelength of the triggered stop-and-go waves. 
If the above instability condition is not fulfilled, instead of
TSG we find a single moving localized cluster (MLC) 
[Figs.~\ref{rho3d}(d) and \ref{rhoQ}(d)], similar to the transition
between stop-and-go waves and localized clusters in homogeneous traffic
\cite{kk-94}. The respective 
boundary is given by
\be
\label{SLC-TSG}
\mbox{TSG-MLC:}\ \  Q_{\rm rmp}=Q_{\rm c2}-Q_{\rm in} \, .
\ee
The MLC continues the course of the initial perturbation which, however, 
changes its propagation speed at the ramp, probably 
because of $\tilde{Q}_{\rm out} < Q_{\rm out}$. 
If $Q_{\rm in} < Q_{\rm c1}$,
the upstream flow cannot sustain localized clusters, since it
is stable to perturbations, and the MLC
becomes a pinned (standing) localized cluster (PLC [or SLC]) \cite{kerner-ramp}
at the on-ramp [Fig.~\ref{rho3d}(e)]:
\be
\label{SW-SLC}
\mbox{MLC-PLC:}\ \  Q_{\rm in} = Q_{\rm c1} \, .
\ee
\par
A pinned localized cluster will grow and, thereby, give
rise to a an extended form of congested traffic (CT), if
the sum $(Q_{\rm in} + Q_{\rm rmp})$ of the inflows 
exceeds the self-organized outflow $\tilde{Q}_{\rm out}$ 
from CT behind ramps.
The corresponding boundaries are given by 
\be
\label{SW-ST}
\mbox{PLC-CT:}\ \  Q_{\rm rmp} = \tilde{Q}_{\rm out} - Q_{\rm in}\, .
\ee
They are the same for HCT and OCT, so that there could be also stationary
and oscillating variants of PLC (SPLC and OPLC).
\par
Finally, 
no congested state can be formed, if
$Q_{\rm down} < Q_{\rm c1}$, since traffic flow is stable to any
perturbation, then.
This defines the boundary between the pinned localized cluster
state and free traffic (FT):
\be
\label{FT-SW}
\mbox{PLC-FT:}\ \  Q_{\rm rmp} = Q_{\rm c1}- Q_{\rm in} \, .
\ee
Due to the metastability between $\rho_{\rm c1}$ and $\rho_{\rm c2}$,
the transition to free traffic depends on the amplitude of the perturbation.
For small perturbation amplitudes, stable traffic flow associated with
free traffic expands up to $Q_{\rm down} < Q_{\rm c2}$. Hence, the
MLC and PLC regimes may disappear in cases of small disturbances, leading 
to the boundaries FT-TSG, FT-OCT, and FT-HCT. In conclusion,
the transition between free and congested traffic 
is of first order (i.e. hysteretic) \cite{Letter}, while the other transitions
seem to be continuous.
\par
Notice that the triple points A(TSG-MLC-PLC), B(OCT-TSG-PLC),
and C(HCT-OCT-PLC) must each satisfy the conditions of 
{\em two} boundaries. In particular, points A
and C are determined uniquely by dynamic properties
on a homogeneous road {\it without} ramps or other inhomogeneities:
$Q^A_{\rm in} = Q_{\rm c1}$, 
$Q^A_{\rm rmp} = Q_{\rm c2} - Q_{\rm c1}$, 
$Q^C_{\rm in} = Q_{\rm cv}$,
and
$Q^C_{\rm rmp} = \tilde{Q}_{\rm out} - Q_{\rm cv} 
\approx Q_{\rm c2} - Q_{\rm cv}$.
\par
Summarizing our results, we have presented a phase diagram of 
traffic states developing close to an on-ramp when passed by a
perturbation on the freeway. 
Similar results are found for disturbances 
on the ramp (not displayed). Moreover, 
the same phase diagram is expected for
other kinds of inhomogeneities as caused by gradients, 
changes in the number of lanes, etc.
In such cases, 
$Q_{\rm rmp}$ corresponds
to the capacity drop along the inhomogeneity.
Despite of the complex behavior of
the model, it was possible to determine the phase boundaries 
analytically and in good agreement with numerical investigations.
As the analytical relations for the phase boundaries do not contain any
details of the nonlocal, gas-kinetic traffic model 
used for the simulations, they are expected to be
also valid for other 
microscopic and macroscopic traffic models which have 
(a) the same instability diagram or, more exactly,
critical densities $\rho_{\rm c1}$, $\rho_{\rm c2}$, and $\rho_{\rm cv}$,
and (b) a characteristic outflow $\tilde{Q}_{\rm out}(Q_{\rm rmp},L_{\rm rmp})$ 
that is independent of the type of congestion at the ramp. A 
sensitive dependence of $\tilde{Q}_{\rm out}$ on the 
surrounding traffic situation 
could lead to different results. 
\par
There is some evidence \cite{Letter,LettA}
that the homogeneous congested state can be identified with the observed
synchronized traffic of type (i) according to the classification
by Kerner and Rehborn \cite{kr-96b}, 
whereas synchronized traffic of types (ii) and (iii) probably
corresponds to the oscillating congested state. Stop-and-go waves
and localized clusters have also their empirical
counterparts \cite{kr-96b,kr-96a}. 
Our findings suggest that inhomogeneities may be the
generic reason for the formation of the different congested states
\cite{Daganzo}. The dependence of the forming states 
on the inflows $Q_{\rm in}$ and $Q_{\rm rmp}$ 
allows to understand the observed transitions
between localized clusters, stop-and-go waves, free and synchronized 
flow \cite{kr-96b,kr-97,kr-96a,emp}. 
It is also of importance for designing on-ramp controls.
The dependence of the traffic states on the ramp length $L_{\rm rmp}$ 
(because of the 
dependence of $\tilde{Q}_{\rm out}$ on $L_{\rm rmp}$) is
relevant for an optimal dimensioning of ramps and explains why the
observable traffic states are dependent on the observation site.
\par
The authors want to thank for financial support by the BMBF (research
project SANDY, grant No.~13N7092) and by the DFG (Heisenberg scholarship
He 2789/1-1).


\begin{references}
\bibitem{TGF95} D. E. Wolf {\em et al.},
{\em Traffic and Granular Flow} (World Scientific, Singapore, 1996);
M. Schreckenberg and D. E. Wolf 
(eds.) {\em Traffic and Granular Flow '97} 
(Springer, Singapore, 1998).

\bibitem{book} 
D. Helbing, {\em Verkehrsdynamik} (Springer, Berlin, 1997).

\bibitem{Schreck} K. Nagel and M. Schreckenberg, 
J. Phys. I France {\bf 2}, 2221 (1992).

\bibitem{kk-94} B.~S.~Kerner and P.~Konh\"auser,
Phys. Rev. E {\bf 50}, 54 (1994).

\bibitem{Letter} D. Helbing and M. Treiber, Phys. Rev. Lett. {\bf 81}, 3042
(1998).

\bibitem{kr-96b} B.~S.~Kerner and H.~Rehborn, 
{\rm Phys. Rev. E} {\bf 53}, R4275 (1996).

\bibitem{kr-97}
B.~S.~Kerner and H.~Rehborn, Phys. Rev. Lett. {\bf 79}, 4030 (1997).

\bibitem{Lee} H. Y. Lee {\em et al.},
Phys. Rev. Lett. {\bf 81}, 1130 (1998).

\bibitem{phase}
J. Krug, Phys. Rev. Lett. {\bf 67}, 1882 (1991).

\bibitem{sym}
M. R. Evans {\em et al.},
Phys. Rev. Lett. {\bf 74}, 208 (1995).
%
%
%

\bibitem{manneville}
P. Manneville,
    {\it Dissipative Structures and Weak Turbulence}
    (Academic Press, New York, 1990).

\bibitem{Collet} J.-P. Eckmann {\em et al.},
Annales de l'Institut Henri Poincare Physique Theorique {\bf 58},
287 (1993);
Y. Kuramoto, Progress in Theor. Phys. Suppl. {\bf 99},
244 (1989).

\bibitem{overview} M. Treiber {\em et al.},
Phys. Rev. E {\bf 59}, 239 (1999).

\bibitem{kr-96a} B.~S.~Kerner and H.~Rehborn, 
{\rm Phys. Rev. E} {\bf 53}, R1297 (1996). 


\bibitem{kerner-ramp} B.~S.~Kerner {\em et al.},
Phys. Rev. E {\bf 51}, 6243 (1995).

\bibitem{LettA} M. Treiber and D. Helbing,
J. Phys. A: Math. Gen. {\bf 32}, L17 (1999).

\bibitem{Daganzo} C. F. Daganzo {\em et al.}, 
{Transpn. Res. B}, in print (1999).

\bibitem{emp} D. Helbing, Phys. Rev. E {\bf 55}, R25 (1997).

\end{references}



\unitlength1mm 

\newpage

\unitlength1mm 


\newpage

\unitlength1.6mm 

\end{document}